\documentclass[12pt]{iopart}
\usepackage{iopams}
\usepackage{natbib}
\usepackage{amsmath,amsfonts,amssymb}
\usepackage{graphicx}
\usepackage{braket}
\usepackage{float}
\usepackage[pdftex]{hyperref} 
\hypersetup{
	colorlinks = true,
	linkcolor = blue,
	anchorcolor = blue,
	citecolor = blue,
	filecolor = blue,
	urlcolor = blue}

\usepackage{tikz}

\begin{document}

\title[Thermodynamic decoupling in the deep-strong coupling regime]{Thermodynamic decoupling in the deep-strong coupling regime}

\author{S. Palafox$^{1}$, M. Salado-Mej\'ia$^{1}$, M. Santiago-Garc\'ia$^{1}$ and R. Rom\'an-Ancheyta$^{2 *}$}

\address{$^{1}$ Instituto Nacional de Astrof\'isica, \'Optica y Electr\'onica, Luis Enrique Erro 1, Santa Mar\'ia Tonantzintla, Puebla, 72840 Mexico}

\address{$^{2}$ Centro de F\'isica Aplicada y Tecnolog\'ia Avanzada, Universidad Nacional Aut\'onoma de M\'exico, Boulevard Juriquilla 3001, Quer\'etaro 76230, Mexico}
\ead{$^*$ancheyta@fata.unam.mx}


\begin{abstract}
In the deep-strong coupling (DSC) regime, the interaction between light and matter exceeds their bare frequencies, leading to an effective decoupling. Theoretical and experimental evidence for this behavior has relied solely on measurements of local observables at equilibrium. However, such a local approach is insufficient to accurately describe energy fluxes in critical and nonequilibrium phenomena. Here, we use a two-terminal quantum junction to derive a thermodynamically consistent global master equation. We demonstrate that the associated heat current, a key nonlocal observable in any quantum thermal machine, also approaches zero in this extreme coupling scenario, underscoring the role of virtual photons in the vacuum ground state. Our results indicate that the decoupling is a more general feature of the DSC regime, with implications for quantum thermotronics. 

\end{abstract}

\section{Introduction}

The effective decoupling is a counterintuitive quantum-optical phenomenon in which the Purcell enhancement of radiative damping, expected to grow quadratically with increasing light-matter coupling, instead collapses in the DSC regime~\cite{DeLiberato,Ripoll2015,ImammogluPRL21}. This arises because material dipoles expel the electric field, dressing electronic states with a significant population of virtual photons, a population that remains largely unaffected even in loss-dominated systems~\cite{Liberato17}. While previous experimental evidence for this decoupling has focused on local observables at equilibrium~\cite{Bayer2017TerahertzLI,StephanieReich,Scalari25arX,Kono25arX}, the present work theoretically investigates whether similar behavior emerges for nonlocal observables under nonequilibrium conditions.

The continuous increase in the light-matter coupling strength $g$~\cite{ReviewNori,FornRMP19} has led to record values~\cite{LangeNatComm24} of the Rabi splitting between hybrid light-matter quasiparticles known as polaritons~\cite{Asenjo_2021}. This trend is driven by both fundamental physics and practical applications~\cite{QIN20241}. For instance, the DSC regime enables more efficient interactions~\cite{ReviewNori}, nonlinear optics without photons~\cite{SavastaPRA17}, ultrafast quantum gates~\cite{RomeroPRL12}, altered chemical reaction rates~\cite{CiutiScience21}, and enhanced thermometric sensors~\cite{Salado21}, among other advantages. The DSC regime is also paramount for vacuum bulk material engineering~\cite{KonoOptics25,Reich25arXiv}. However, despite the abundance of its cavity-free polaritons~\cite{Canales21}, such as those in water droplets from clouds~\cite{CanalesPRL24}, relatively few studies have examined the impact of such large couplings on heat transport \textcolor{black}{using either the quantum Rabi model~\cite{PaladinoQST25,Paladino25arXiv,YamamotoPRB25} or a single spin degree of freedom~\cite{SegalPRB25}. Recently, new hyperbolic media under non-Hermitian conditions have experimentally demonstrated asymmetric polaritonic transport~\cite{GUO20243491}.}

In recent room-temperature experiments reaching the DSC regime, it is crucial to include the counterrotating and diamagnetic terms of the Hopfield Hamiltonian to accurately reproduce transmission spectra~\cite{Kono25arX, LangeNatComm24}. Incorporating these non-energy-preserving contributions, however, requires adopting a global (microscopic) framework~\cite{Haack17,Gonzalez18JPA} to analyze the nonequilibrium DSC regime. In contrast to the local (phenomenological) approach, which does not adequately characterize intersystem correlations~\cite{EricLutz}, the global description yields a thermodynamically consistent master equation~\cite{Tahir18,Potts_2021}, thereby guaranteeing compliance with the second law~\cite{Kosloff,wideband-limit}.

\begin{figure}[b]
\centering
\includegraphics[scale=0.35]{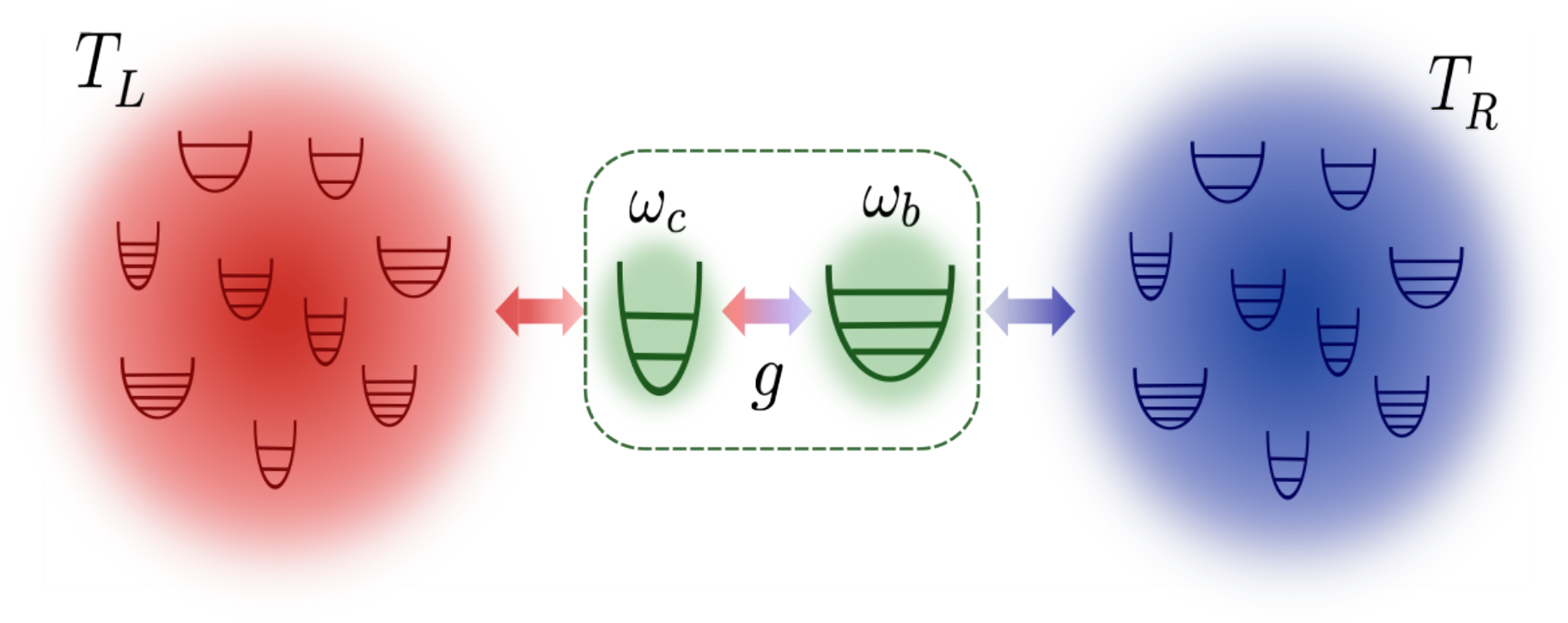}
\caption{Schematic representation of a thermal junction model. The central system (enclosed by the dashed green rectangle) consists of two coupled quantum oscillators with bare frequencies, $\omega_{c}^{}$ and $\omega_{b}^{}$. Each oscillator is weakly and locally coupled to a thermal bath—modeled as a collection of quantum harmonic oscillators—at temperatures $T_L^{}>T_R^{}$. The coupling strength between the central subsystems, $g$, may operate in the ultrastrong coupling or deep-strong coupling regime.}
\label{system}
\end{figure}

In this work, we introduce the concept of {\it thermodynamic decoupling} to describe the suppression of the heat current, a nonlocal observable, between two thermal baths connected by two coupled systems in the DSC regime, see Fig. \ref{system}. This situation resembles that of an atomic junction~\cite{Palafox_PRE_2022}. We derive a Lindblad global master equation for the model and provide analytical expressions for the dissipation rates of the upper and lower polaritons, which were not previously reported. These rates are temperature-independent, demonstrating the robustness of the breakdown of the Purcell effect and offering insights into their experimental observations at room temperature.

\textcolor{black}{In contrast to~\cite{DeLiberato,Ripoll2015,ImammogluPRL21,Liberato17}, w}e calculate the steady-state heat current and derive compact expressions for the population of virtual photons across the full spectrum, from the weak coupling limit to the DSC regime. Our results show that, in the DSC regime, the heat current vanishes as $g^{-1}$, irrespective of resonant conditions. This suppression occurs as the virtual photons grow linearly with $g$, a feature expected to be accessible in forthcoming quantum heat transport experiments~\cite{Pekola24arXv}. Our results may have implications in the field of quantum thermodynamics~\cite{CampellRoadMap}, particularly in the design of quantum thermal machines~\cite{Nicole_2024}, where precise computation and control of the heat currents are essential~\cite{MoroniQST,Nicole25NatPhy}. 

\section{The model}\label{sec:ME}

We begin by considering the simplest two-mode version of the Hopfield Hamiltonian~\cite{ReviewNori}
\begin{equation}\label{Hopfield}
H_S^{}=\omega_c^{} a_L^\dagger a_L^{} +\omega_b^{} a_R^\dagger a_R^{}+ig_1^{}(a_L^{}a_R^\dagger-a_L^\dagger a_R^{})+ig_2^{}(a_L^\dagger a_R^\dagger-a_L^{}a_R^{})+D(a_L^{}+a_L^\dagger)^2, 
\end{equation}
where we set $\hbar=1$. Here, ($a_L^{},a_L^\dagger$) and ($a_R^{},a_R^\dagger$) are, respectively, the annihilation and creation operators of the photonic field (left central oscillator in Fig.~\ref{system}) with frequency $\omega_c^{}$, and the collective material excitations (right oscillator) with frequency $\omega_b^{}$. Since the matter part is often an ensemble made of a large number of two-level systems, we have assumed that one can bosonize them~\cite{LiberatoJOSAB24}. Therefore, the commutation relations are $[a_\lambda^{},a_\lambda^\dagger]=1$, where $\lambda=L, R$. The corotating (counterrotating) coupling strength is $g_{1}^{}$ ($g_2^{}$), and $D$ corresponds to the diamagnetic term, known as the self-interaction energy. \textcolor{black}{Note that the $D$ term was not considered in~\cite{PaladinoQST25,Paladino25arXiv,YamamotoPRB25,SegalPRB25}.}

Although it is experimentally feasible to engineer anisotropic coupling ($g_1^{}\neq g_2^{}$)~\cite{Yu24NatComm}, for simplicity, we restrict our analysis to the isotropic case $g_1^{}=g_2^{}\equiv g$. The DSC regime is defined by $g/ \omega_{c,b}^{}>1$, while the ultrastrong coupling (USC) corresponds to $0.1\leq g/\omega_{c,b}^{}\leq1$~\cite{ReviewNori,FornRMP19}. Both regimes go beyond the standard rotating-wave approximation (RWA). To derive the master equation of this coupled system within the global (microscopic) approach~\cite{Haack17,Gonzalez18JPA}, it is first necessary to diagonalize $H_S^{}$. We apply two simple yet non-trivial unitary transformations, $T=\exp(-i\pi a_R^\dagger a_R^{}/2)$,  $R_\theta^{}=\exp[i\theta (x p_y^{}-y p_x^{})]$, such that $R_\theta TH_S^{}T^\dagger R_\theta^\dagger\equiv H_{\rm diag}$, which yields the diagonal Hamiltonian $H_{\rm diag}=\sum_{j\in \{x,y\}}\frac{1}{2}(p_j^2+\omega_j^2j^2)$, as shown in~\cite{Salado21}. Here, $\omega_x^{}$ ($\omega_y^{}$) is the frequency of the upper (lower) polariton
\begin{equation}\label{w_xy}
2\,\omega_{x,y}^{\,2}=(\omega_c^2+\omega_b^2+4D\omega_c^{})\pm\big[(\omega_c^2-\omega_b^2+4D\omega_c^{})^2+16 g^2\omega_c^{}\omega_b^{}\big]^\frac{1}{2}.
\end{equation}
 Since $p_{x,y}^{}$ ($x,y$) are the canonical momentum (position) Hermitian operators, i.e., $[x,p_x^{}]=i$, then $H_{\rm diag}^{}$ describes two uncoupled quantum harmonic oscillators with eigenvalues $E_{mn}=\omega_x(m+1/2)+\omega_y(n+1/2)$, where $n,m\in\mathbb{Z}^{+}$. $R_\theta^{}$ represents a rotation by an angle $\theta$ which, for $\omega_c^{}>\omega_b^{}$, it can be obtained from $\tan(2\theta)=4g(\omega_b^{}\omega_c^{})^{\frac{1}{2}}/(\omega_c^2+4D\omega_c^{}-\omega_b^2)$. As we will consider the Thomas-Reiche-Kuhn (TRK) sum rule in light-matter interaction, $D$ takes the value $D=g^2/\omega_b^{}$~\cite{Wang1999}. However, this means that if $\omega_c^{}<\omega_b^{}$, the denominator in $\theta$ becomes problematic when $g$ varies from the weak-coupling to the DSC regime. Therefore, to maintain the continuity of $R_\theta^{}$ in that situation, one needs to use the angle $\theta'\equiv\theta+\pi/2$ for $g\leq g_{\rm crit}^{}$, and $\theta'=\theta$ for $g>g_{\rm crit}^{}$, where  $g_{\rm crit}^2\equiv\omega_b^{}(\omega_b^2-\omega_c^2)/4\omega_c^{}$. Since the eigenvalues $E_{mn}$ are well defined for any value of $g$, the replacement by the angle $\theta'$ constitutes a subtle yet crucial condition that was overlooked in~\cite{Salado21}. It is also instructive to note that, in the uncoupled limit, one finds $\lim_{g\rightarrow0}^{}\omega_{x,y}^{}=\omega_{c,b}^{}$ ($\omega_{b,c}^{}$) when  $\omega_c^{}>\omega_b^{}$ ($\omega_b^{}>\omega_c^{}$). For the remainder of the article, we will adopt the former case, without loss of generality.

The system Hamiltonian in terms of its eigenoperators is $ {H}_{\rm diag}^{}=\omega_x^{}(A_L^\dagger A_L^{}+\frac{1}{2})+\omega_y^{}(A_R^\dagger A_R^{}+\frac{1}{2})$,
where we have defined $A_L^{}=(\omega_x^{} x+ip_x^{})/\sqrt{2\omega_x^{}}$ and $A_R^{}=(\omega_y^{} y+ip_y^{})/\sqrt{2\omega_y^{}}$, which correspond to the annihilation operators of the upper and lower polaritons, respectively. In the uncoupled limit $A_\lambda^{}\rightarrow a_\lambda^{}$.  These satisfy the commutation relations $[A_\lambda^{},A_\lambda^\dagger]=1$, $[{H}_{\rm diag}^{},A_{L}^{}]=-\omega_{x}^{} A_{L}^{}$ and $[{H}_{\rm diag}^{},A_R^{}]=-\omega_y^{} A_R^{}$. 
The relations between the bare ($a_\lambda^{}$, $a_{\lambda}^\dagger$) and dressed ($A_\lambda^{}$, $A_{\lambda}^\dagger$) operators are
\begin{subequations}\label{RelationsOp}
\begin{align}
T_D^{} a_L^{} T_D^\dagger=&\,( f_1^{} A_L^{}+f_2^{} A_L^\dagger-f_3^{} A_R^{}-f_4^{} A_R^\dagger),\label{aLandA} 
\\ 
T_D^{} a_R^{} T_D^\dagger=& i(f_5^{} A_L^{}+f_6^{} A_L^\dagger+f_7^{} A_R^{}+f_8^{} A_R^\dagger),\label{aRandAR}
\end{align}
\end{subequations}
where $T_D^{}\equiv R_\theta^{} T$. The coefficients $f_i^{}$ are 
\begin{equation}\label{fCoef}
f_{1,2}^{}=\frac{\omega_c^{} \pm \omega_x^{}}{2\sqrt{\omega_x^{}\omega_c^{}}}\cos\theta,\,\,\,
f_{3,4}^{}=\frac{\omega_c^{} \pm \omega_y^{}}{2\sqrt{\omega_y^{}\omega_c^{}}}\sin\theta,\,\,\,
 f_{5,6}^{}=\frac{\omega_b^{} \pm \omega_x^{}}{2\sqrt{\omega_x^{}\omega_b^{}}}\sin\theta,\,\, \,
f_{7,8}^{}=\frac{\omega_b^{} \pm \omega_y^{}}{2\sqrt{\omega_y^{}\omega_b^{}}}\cos\theta.
\end{equation}
We follow the well-established derivation~\cite{Carmichael-Statistical_Methods_1, Breuer07, Rivas2011} to obtain, under the Born-Markov approximation, the master equation for the  density operator $\rho$ of the central system~\cite{Zhou2020}
\begin{equation}\label{rho}
\frac{d\rho}{dt}=-i{\omega}_x^{}[A_L^{\dagger}A_L^{},\rho]-i{\omega}_y^{}[A_R^{\dagger}A_R^{},\rho]+\alpha_1^{}\mathcal{L}[A_L^{}]\rho+\beta_1^{}\mathcal{L}[A_R^{}]\rho+\alpha_2^{}\mathcal{L}[A_L^{\dagger}]\rho+\beta_2^{}\mathcal{L}[A_R^{\dagger}]\rho,
\end{equation}
where $\mathcal{L}[\mathcal{O}]\rho\equiv 2\mathcal{O}\rho \mathcal{O}^\dagger-\mathcal{O}^\dagger \mathcal{O}\rho-\rho \mathcal{O}^\dagger \mathcal{O}$ is a global Lindblad super-operator and
\begin{equation}\label{dissipationRates}
    \begin{split}
    \alpha_{1,2}^{}\!=\!\gamma_{_{L}}\!(\omega_x^{})(f_1^{}\!+\!f_2^{})^2\big[n(\omega_x^{},T_L^{})\!+\!(1\!\pm\!1)/2\big]
\!+\!\gamma_{_{R}}\!(\omega_x^{})(f_5^{}\!-\!f_6^{})^2\big[n(\omega_x^{},T_R^{})\!+\!(1\!\pm\!1)/2\big],\!\! 
\\
\beta_{1,2}^{}\!=\!\gamma_{_{L}}\!(\omega_y^{})(f_3^{}\!+\!f_4^{})^2\big[n(\omega_y^{},T_L^{})\!+\!(1\!\pm\!1)/2\big]
\!+\!\gamma_{_{R}}\!(\omega_y^{})(f_7^{}\!-\!f_8^{})^2\big[n(\omega_y^{},T_R^{})\!+\!(1\!\pm\!1)/2\big].\!\!
    \end{split}
\end{equation}
The Bose-Einstein distribution is $n(\omega_{\!j}^{},\!T_\lambda^{})\equiv[\exp({\hbar \omega_{\!j}^{}/k_B^{} T_\lambda^{}})-1]^{-1}$, $T_\lambda^{}$ is the temperature of the bath $\lambda$, $k_B^{}$ is the Boltzmann constant, $\gamma_\lambda^{}(\omega_{x,y}^{})=\pi \sigma(\omega_{x,y}^{})|g_\lambda^{}(\omega_{x,y}^{})|^2$ is the coupling strength to the bath $\lambda$, $\sigma(\omega_{x,y}^{})$ is density of states, and $|g_\lambda^{}(\omega_{x,y}^{})|^2$ comes from the local system-bath weak interaction~\cite{Carmichael-Statistical_Methods_1}. \textcolor{black}{We emphasize that in this work, the system-bath coupling remains weak and is quantified by $\gamma_\lambda$, regardless of the fact that the normalized coupling ($g/\omega_{c,b}$) can take large values. With this, the use of a local master equation (LME) is justified when $g \lesssim \gamma_\lambda \ll \omega_{c,b}$~\cite{NJP_Geraldine_2020}, while the global master equation (GME) holds when $g \gg \gamma_\lambda^{}$ or $|\omega_c - \omega_b| \gg \gamma_\lambda$~\cite{Cattaneo_2019}. For a detailed comparison between the LME and the GME in the context of quantum thermal machines, see~\cite{Haack17}}. As we will in Sec.~\ref{Heat-current}, it is useful to rewrite Eq.~(\ref{rho}) as $\dot\rho=-i[H_{\rm diag}^{},\rho]+\mathcal{L}_{\!L}^{}\rho+\mathcal{L}_{\!R}^{}\rho$, where $\mathcal{L}_{\!\lambda}^{}\rho$ is defined in (\ref{L_Lambda}).

To obtain the expectation value $\langle\mathcal{O} \rangle\equiv{\rm tr}\{\rho\,\mathcal{O}\}$ of an arbitrary operator $\mathcal{O}$, we use Eq.~(\ref{rho}) and get the corresponding differential equation of motion $d\langle \mathcal{O}\rangle/dt={\rm tr}\{\dot\rho\,\mathcal{O}\}$. For the mean excitation number of each polariton, the time-dependent solutions are
\begin{equation}\label{time_solutions}
    \langle A_L^\dagger A_L^{}\rangle=\frac{\alpha_2^{}}{\alpha_1^{}-\alpha_2^{}}\big[1-e^{-2(\alpha_1^{}-\alpha_2^{})t}\,\big],
    \quad
    \langle A_R^\dagger A_R^{}\rangle=\frac{\beta_2^{}}{\beta_1^{}-\beta_2^{}}\big[1-e^{-2(\beta_1^{}-\beta_2^{})t}\,\big].
\end{equation}
From these, we identify two temperature-independent decay rates, which we define as
\begin{subequations}\label{decayrates}
\begin{eqnarray}
    \Gamma_{\!x}&\equiv &\alpha_1^{}-\alpha_2^{}=\gamma_{_L}^{}\!(\omega_x^{})(f_1^{}+f_2^{})^2+\gamma_{_R}^{}\!(\omega_x^{})(f_5^{}-f_6^{})^2,
    \\
    \Gamma_{\!y}&\equiv &\beta_1^{}-\beta_2^{}=\,\gamma_{_L}^{}\!(\omega_y^{})(f_3^{}+f_4^{})^2+\gamma_{_R}^{}\!(\omega_y^{})(f_7^{}-f_8^{})^2.
\end{eqnarray}
\end{subequations}

\textcolor{black}{
Given the quadratic nature of $H_S$, an explicit expression for $\rho$ could potentially be derived using the superoperator approach~\cite{MOYACESSA20061}. However, this method is algebraically complex and yields cumbersome expressions, even for a single reservoir. We have therefore chosen to utilize the equations of motion for the relevant observables, which allow us to characterize the system's state by computing higher-order moments and correlation functions through the quantum regression formula.}

\section{Breakdown of the Purcell effect}\label{PurcellEffect}

\begin{figure}[t]
\centering
\includegraphics[scale=0.6]{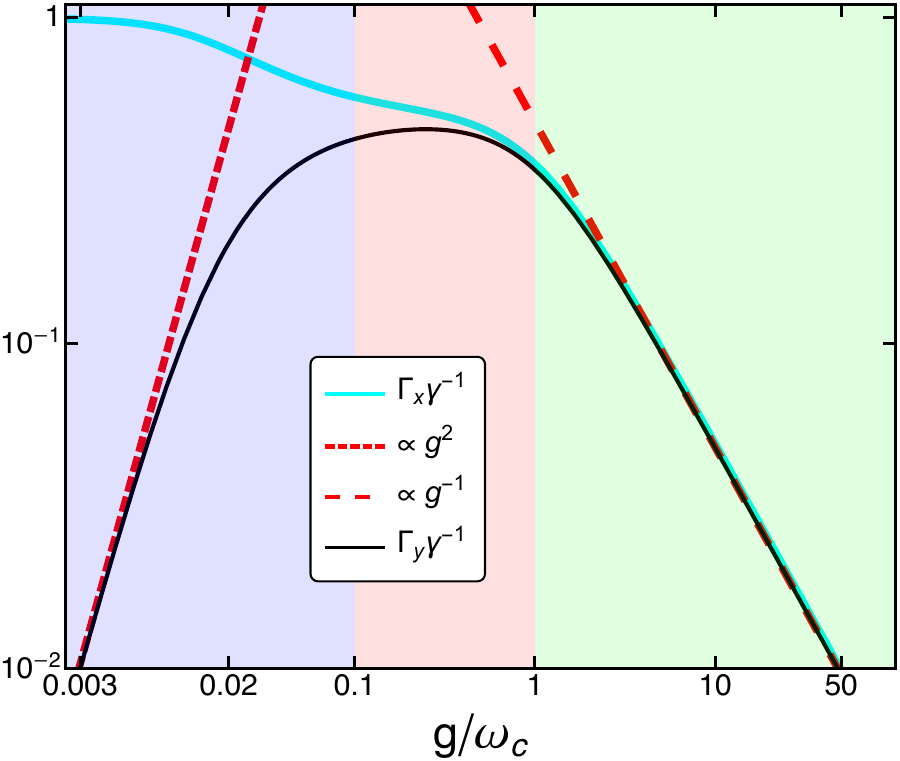}
\caption{Decay rates $\Gamma_{\!x,y}$  of the upper (cyan solid line) and lower (black solid line)  polariton as a function of normalized coupling $g/\omega_c^{}$, and $\omega_b^{}=0.97\omega_c$. See Sec.~\ref{PurcellEffect} for details.}
\label{PurcellEffectPlot}
\end{figure}

To illustrate the breakdown of the Purcell effect in our system more clearly, we assume a flat spectral density, commonly known as the wideband limit~\cite{wideband-limit}, in which the coupling to the bath $\lambda$ is taken as an energy-independent constant value $\gamma_{\lambda}^{}(\omega_{x,y}^{})\equiv\gamma_\lambda^{}$. Then, we disconnect the right bath ($\gamma_{_R}=0$), such that Eqs.~(\ref{decayrates}) reduce to $\Gamma_{\!x}=\gamma_{_L}[\cos^2\theta\,{\omega_c^{}}{\omega_x^{-1}}]$ and $\Gamma_{\!y}=\gamma_{_L}[\sin^2\theta\,{\omega_c^{}}{\omega_y^{-1}}]$. Since the diagonalization angle $\theta$ depends on $g$, $\Gamma_{\!x,y}^{}$ has different behavior depending on the light-matter coupling regime. For instance, in the limit case $g\rightarrow0$, $\Gamma_{\!x}=\gamma_{_L}^{}$ and $\Gamma_{\!y}=0$. This means that $\Gamma_{\!x}$, which only contains the photonic component in this limit, decays independently of the matter system, while the latter is totally isolated. As $g$ increases (for $g/\omega_{b,c}\ll1$), the upper polariton decay rate remains nearly constant at $\Gamma_x\approx\gamma_{_L}$. In contrast, the lower polariton decay rate grows quadratically as $\Gamma_{\!y}\approx4g^2\omega_c^2\gamma_{_L}/(\omega_c^2-\omega_b^2)^2$. This behavior, shown by the short-dashed line in Fig.~\ref{PurcellEffectPlot}, is the signature of the Purcell effect: a radiative process originating from {local} light-matter interactions. While in the USC regime (pink region) $\Gamma_{\!x,y}$ reach a saturation value, they plummets to zero in the DSC (green region) as $\Gamma_{\!x,y}\approx g^{-1}\omega_b^2\gamma_{_L}({\omega_c^{}\omega_b^{}})^\frac{1}{2}(\omega_c^2+\omega_b^2)^{-1}$ (long-dashed line). This behavior, referred to as the breakdown of the Purcell effect, was predicted through an input-output theory~\cite{DeLiberato} and experimentally verified by measuring a decrease in the linewidth of the absorption spectra~\cite{StephanieReich}. To the best of our knowledge, this is the first derivation of this effect using a Lindblad master equation, and also with the simplest two-mode version of the Hopfield Hamiltonian. 

\textcolor{black}{
The intuitive explanation for the effective decoupling is due to the suppression of the electric field near the material dipoles~\cite{DeLiberato,Scalari25arX}. However, our results can be explained simply by the energetic structure of the coupled system. For instance, in the DSC regime, the anharmonic energy spectrum $E_{mn}$ of Sec.~\ref{sec:ME} becomes equispaced, as shown in~\cite{Salado21}. This occurs because, for $g/\omega_c \gg 1$, the lower polariton frequency $\omega_y \approx 0$, while the upper polariton frequency approaches $\omega_x \approx 2g(\omega_c/\omega_b)^{1/2}$. Consequently, in the DSC regime, the energy structure of the coupled system closely resembles that of a single effective harmonic oscillator, indicating effective decoupling.}

\section{Heat current}\label{Heat-current}

The total heat current consists of contributions from both the left and right baths given by~\cite{Haack17,Palafox_PRE_2022} $\mathcal J=\mathcal J_{\!L}^{}+\mathcal J_{\!R}^{}$, where $\mathcal{J}_\lambda^{}={\rm tr}\{H_{\rm diag}^{}\,\mathcal{L_{\!\lambda}^{}\rho}\}$, and $\lambda=L, R$. In the steady state, $\dot\rho_{\!_{\rm SS}}^{}=0$, $\mathcal{J}_{\rm SS}^{}=0$, and $\mathcal{J}_{L}^{\rm SS}=-\mathcal{J}_{R}^{\rm SS}$. This is a consequence of the first law and tells us that energy is conserved. 
The heat current from the left bath is (see \ref{Appendix_1}):
\begin{equation}\label{HeatCurrent}
\mathcal{J}_L^{\rm SS}=\sum_{j\in\{x,y\}}\frac{2\omega_{\!j}^{}\gamma_{_{L}}\!(\omega_j^{})\gamma_{_{R}}\!(\omega_j^{})}
{\gamma_{_{L}}\!(\omega_j^{})\frac{\omega_b^{}}{\omega_j}\sec^2 \theta\!_j^{}+\gamma_{_{R}}\!(\omega_j^{})\frac{\omega_j^{}}{\omega_c^{}}\csc^2 \theta\!_j^{}}\big[n(\omega_j^{},\!T_{\!L}^{})-n(\omega_j^{} ,\!T_{\!R}^{})\big],
\end{equation}
%
%
where $\theta\!_{x,y}^{}\equiv\theta+(1\!\pm\!1)\pi/4$, and we used Eq.~(\ref{fCoef}) to simplify some terms. We introduce the peculiar notation for $\theta_{x,y}$ to obtain a more compact expression for $\mathcal{J}_L^{\rm SS}$. For example, trigonometric identities give $\csc(\theta+\pi/2)\!=\!\sec\theta$ and $\sec(\theta+\pi/2)=-\csc\theta$. An analogous motivation underlies the notation adopted in Eq.~(\ref{dissipationRates}). 
Note that Eq.~(\ref{HeatCurrent}) can be rewritten as $\mathcal{J}_{L}^{\rm SS}={\sum}_{j\in\{x,y\}} \int \mathcal{T}_j^{}(\nu)[n(\nu,T_L^{})-n(\nu,T_R^{})]\nu d\nu$, which is a Landauer-type formula~\cite{Segal_PRL_2005,Pekola_RMP_2021} where
\begin{equation}\label{transm_coeff}
    \mathcal{T}_{\!j}^{}(\nu) =\frac{2\gamma_{_{L}}(\nu)\,\gamma_{_{R}}(\nu)\;\delta (\nu-\omega_{j}^{})}{\gamma_{_{L}}(\nu) \frac{\omega_b^{}}{\nu}\sec^2\theta\!_j^{} + \gamma_{_{R}}(\nu)\frac{\nu}{\omega_c^{}}\csc^2\theta\!_j^{}}
\end{equation}
is the transmission coefficient, and $\delta(\nu-\omega_j^{})$ is the Dirac delta function. It is evident that if any of  $\gamma_\lambda^{}(\omega_{x,y}) = 0$, then $\mathcal{J}_L^{\rm SS}$ must equal 0. Regarding the second law, it is easy to prove that the irreversible entropy production rate, defined by $\Pi=\dot S-\sum_\lambda^{} \mathcal{J}_\lambda^{}/(k_BT_\lambda^{})$~\cite{Landi_RMP_2021}, satisfies $\Pi_{\rm SS}^{}\ge0$. Therefore, heat will always flow from hot to cold baths, making our \textcolor{black}{global} master equation~(\ref{rho}) thermodynamically consistent.

At equilibrium, $T_{\!L}^{}=T_{\!R}^{}\equiv T$, $\mathcal{J}_L^{\rm SS}=0$, and the mean polaritonic populations reduce to $\langle A_L^\dagger A_L^{} \rangle_{\rm SS}^{}=n(\omega_x,T)$ and $\langle A_R^\dagger A_R^{} \rangle_{\rm SS}^{}=n(\omega_y,T)$, which coincide with those of the thermal (Gibbs) state. Consequently, we confirm that $\rho_{\!_{\rm SS}}^{}=\rho_{{\rm th}}^{}=\exp(-\beta H_{\rm diag})/Z$, where $\beta=(k_B^{}T)^{-1}$ and $Z$ is the partition function. Moreover, as the equilibration process depends on $\langle A_\lambda^\dagger A_\lambda^{}\rangle$ [as noted in~(\ref{leftheat})], the time dependence in Eq.~(\ref{time_solutions}) predicts a power-law slowdown of thermalization, which occurs because $\Gamma_{\!x,y}^{}\sim g^{-1}$ in the DSC regime. This behavior contrasts with the exponential slowdown recently predicted for the asymmetric quantum Rabi model~\cite{BernardisPRA23}.



\begin{figure}[t]
\centering
\includegraphics[scale=0.5]{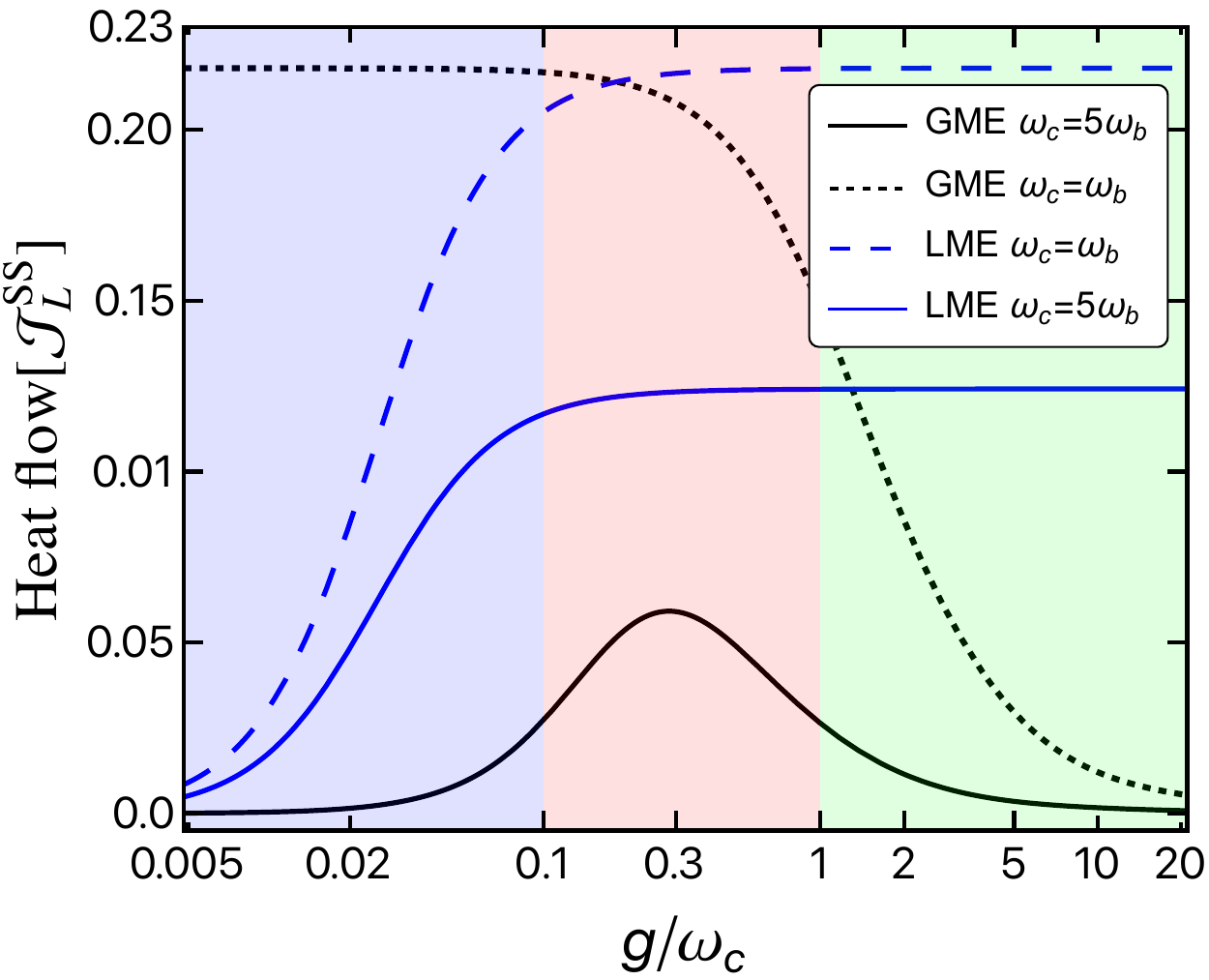}
\caption{Steady-state heat current [Eq.~(\ref{HeatCurrent}) and Eq.~(\ref{LME_RWA})] as a function of normalized coupling $g/\omega_c^{}$ at resonant (dashed line) and nonresonant (solid line) conditions, see Sec.~\ref{Heat-current} for details.}
\label{HeatFlatOhmic}
\end{figure}

To investigate the behavior of the heat current, we evaluate Eq.~(\ref{HeatCurrent}) for both the resonant ($\omega_c^{}=\omega_b^{}$) and non-resonant ($\omega_c^{}=5\omega_b^{}$) conditions. As shown in Fig.~\ref{HeatFlatOhmic}, in the off-resonant case (black solid line), the steady-state heat current $\mathcal{J}_L^{\rm SS}$ initially increases with $g/\omega_c$. However, contrary to previous reports~\cite{Haack17,Tahir18}, as $g/\omega_c$ continues to grow, $\mathcal{J}_L^{\rm SS}$ reaches a maximum and then decreases to zero in the DSC regime. This marks the start of a process we call thermodynamic decoupling, in which heat transport is suppressed as $g^{-1}$. The resonant case (black dashed line) displays similar behavior in the DSC regime; however, for $g/\omega_c \ll 1$, $\mathcal{J}_L^{\rm SS}$ approaches a constant value even as $g\rightarrow 0$. This apparent nonzero heat flow arises from the breakdown of the full secular approximation—essential for deriving the global master equation~(\ref{rho})—in the resonant weak-coupling limit~\cite{Cattaneo_2019}. 

\textcolor{black}{
On the other hand, a LME is valid when $(g/\omega_{c,b}) \lesssim (\gamma_\lambda/\omega_{c,b}) \ll 1$~\cite{NJP_Geraldine_2020}. Thus, we cannot expect accurate physical behavior of the heat current in the DSC regime, which, by definition, lies outside the LME's validity range. The failure of the LME has been demonstrated in various scenarios~\cite{EricLutz}. Nevertheless, it is instructive to examine the predictions of the LME when applied to calculate the heat current in the DSC regime. Within the LME's validity range, we can neglect the counterrotating ($g_2$) and diamagnetic ($D$) terms in Eq.~(\ref{Hopfield}). A derivation of the LME for this simplified Hamiltonian can be found in~\cite{MoroniQST}, which, in our notation and in the steady state, yields the following heat current~\cite{Haack17,MoroniQST}}
\textcolor{black}{
\begin{equation}\label{LME_RWA}
    \mathcal{J}_{L}^\texttt{LME}=\frac{2\omega_c^{}4g^2\gamma_{\!_L}^{}\gamma_{\!_R}^{}}{(\gamma_{\!_L}^{}+\gamma_{\!_R}^{})(\gamma_{\!_L}^{}\gamma_{\!_R}^{}+4g^2)}\big[n(\omega_c^{},\!T_{\!L}^{})-n(\omega_b^{} ,\!T_{\!R}^{})\big].
\end{equation}}
\textcolor{black}{
In Fig.~\ref{HeatFlatOhmic}, we show $\mathcal{J}_{L}^\texttt{LME}$ for the resonant (blue dashed line) and off-resonant (blue solid line) cases. In both scenarios, unlike the predictions of the GME, the heat current remains nonzero in the DSC regime; however, the LME is not justified in this regime. Furthermore, Eq.~(\ref{LME_RWA}) predicts heat flow from the cold reservoir to the hot one whenever $\frac{\omega_b}{T_R^{}} < \frac{\omega_c}{T_L^{}}$, which clearly violates the second law, as first noted in~\cite{Kosloff}.}
To generate Fig.~\ref{HeatFlatOhmic}, we use the parameters $k_{\!B}^{}T_L = 5$, $k_{\!B}^{}T_R = 0.5$, and $\gamma_{_L} = \gamma_{_R} = 0.05\omega_c$, adopted from~\cite{Haack17}.

\section{Virtual excitations}\label{sec:Virtualphotons}

\begin{figure}[t]
\centering
\includegraphics[scale=0.65]{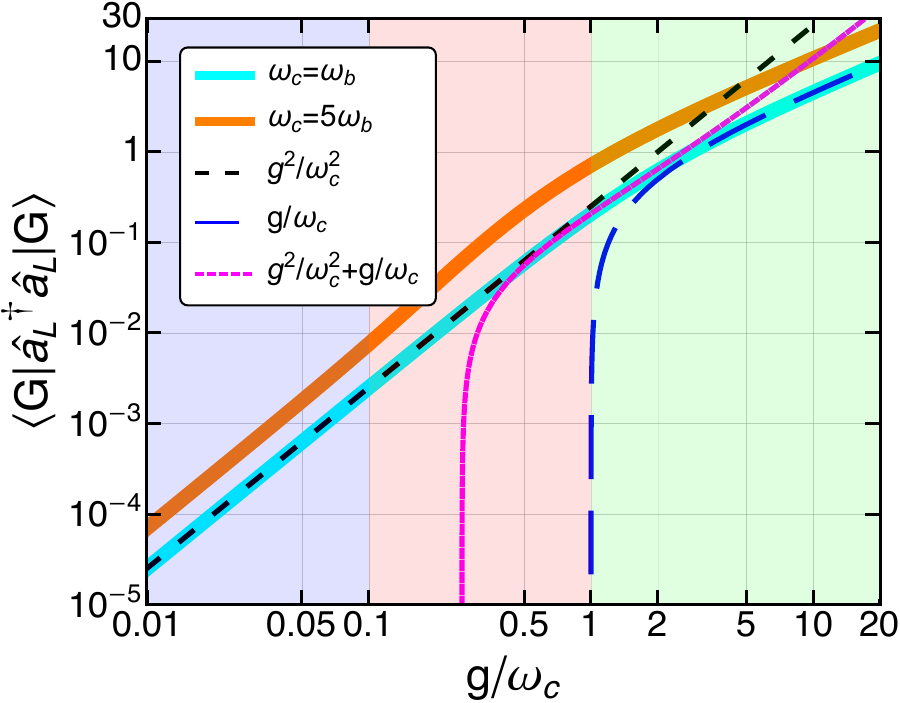}
\caption{Virtual photons population, $\langle a_L^\dagger a_L^{}\rangle_{{\rm SS}}^{}=f_2^2+f_4^2$, both out of resonance (orange solid line) and in resonance (cyan solid line) conditions. Depending on the normalized coupling ($g/\omega_c^{}$): from 0.01(weak) to 0.5(ultra-strong), the virtual photons contribution is well approximated by $(g/\omega_c)^2$ (black dashed line). Conversely, when we are over 2(deep-strong) the contribution is $\propto g/\omega_c^{}$ (blue dashed line), see Eq.~(\ref{virtualphotonsdeep}). From 0.5 (USC) until 2 (DSC), the behavior can be captured by $g^2/\omega^2+g/\omega$ (magenta dashed line), see Eq.~(\ref{virtualphotonsinter}). }
\label{virtualphotonsplot}. 
\end{figure}

The presence of virtual photons is usually negligible when we work in the weak and USC regimes. However, in the DSC regime, these virtual excitations grow to such an extent that it is necessary to consider them~\cite{ReviewNori}. In this section, we derive practical expressions for the population of virtual photons in the ground state of the coupled system. Using Eq.~(\ref{aLandA}), we obtain 
    \begin{align}
    \langle a_L^\dagger a_L^{}\rangle_{{\rm SS}}^{}=\frac{\alpha_2^{}}{\alpha_1^{}-\alpha_2^{}}(f_1^2+f_2^2)+\frac{\beta_2^{}}{\beta_1^{}-\beta_2^{}}(f_3^2+f_4^2)+f_2^2+f_4^2.\label{AverageaL}
\end{align}
This expression is valid for any $T_\lambda^{}$, but at $T_\lambda^{}=0$ the system is in the ground state $|G\rangle\langle G|$, and the coefficients satisfy $\alpha_2^{}=\beta_2^{}=0$. This outcome is something to be expected from a true thermalizing master equation~\cite{circuitQED-Blais}. In this case, the bare excitations becomes $\langle a_L^\dagger a_L^{}\rangle_{{\rm SS}}^{}=f_2^2+f_4^2$. The fact that $\langle a_L^\dagger a_L^{}\rangle_{{\rm SS}}^{}\neq0$ in the ground state indicates the presence of a finite population of virtual photons. Due to the TRK sum rule, the polaritonic and bare frequencies satisfy $\omega_x^{}\omega_y^{}=\omega_c^{}\omega_b^{}$, which in turn implies $f_2^2+f_4^2=f_6^2+f_8^2$, and therefore $\langle a_L^\dagger a_L^{}\rangle_{{\rm SS}}^{}=\langle a_R^\dagger a_R^{}\rangle_{{\rm SS}}^{}$. This result coincides with that reported in~\cite{TufarelliPRA15,ReviewNori}, and in~\cite{EricLutz} for the case $T_\lambda^{}\neq0$ and $D=0$ due to criticality. Under resonance conditions ($\omega_c^{}=\omega_b^{}\equiv\omega$) and $g/\omega \ll 1$, the virtual-photon population reduces to $\langle G|a_L^\dagger a_L|G\rangle\approx\frac{1}{4}\left(\frac{g}{\omega}\right)^2$. In typical cavity-QED experiments, the normalized coupling is $g/\omega\sim10^{-6}$~\cite{FornRMP19,Devorete07}, so that the resulting virtual-photon population, of order $\sim10^{-12}$, is entirely negligible. The same applies to strong-coupling circuit-QED studies with amorphous materials~\cite{Ancheyta_2022}.
For the opposite case, when $g/\omega \gg 1$ we get
\begin{equation}\label{virtualphotonsdeep}
\langle G|a_L^\dagger a_L^{}|G\rangle\approx\frac{1}{2}\left(\frac{g}{\omega}\right)-\frac{1}{2}.
\end{equation}
For the current record value $g/\omega=3.19$~\cite{LangeNatComm24}, Eq.~(\ref{virtualphotonsdeep}) predicts a population of $1.1$ virtual photons with an error below $6\%$. As experiments advance toward stronger light-matter couplings, our Eq.~(\ref{virtualphotonsdeep}) is expected to gain both accuracy and significance.
For the intermediate regime $g/\omega\sim 1$, we obtain
\begin{equation}\label{virtualphotonsinter}
\langle G|a_L^\dagger a_L^{}|G\rangle\approx \frac{1}{6}\left(\frac{g}{\omega}\right)+\frac{1}{11}\left(\frac{g}{\omega}\right)^2-\frac{1}{20}.
\end{equation}
This expression, relevant to several ongoing experiments in the DSC regime, reproduces exactly, for example, the virtual-photon population reported in~\cite{Bayer2017TerahertzLI,Kono25arX}. Finally, in Fig.~\ref{virtualphotonsplot}, we compare the three approximate formulas for the virtual photons, illustrating how each captures the behavior in its respective coupling regime. The cyan (orange) solid line corresponds to the exact value $f_2^2+f_4^2$ at resonant (non-resonant) conditions. 
\textcolor{black}{Note that the orange curve is above the cyan curve, indicating that the coupled system contains more virtual photons under these non-resonant conditions. Notably, the associated heat current (solid black line in Fig.~\ref{HeatFlatOhmic}) reflects this behavior, showing faster suppression (green region) compared to the resonant case (dashed black line in Fig.~\ref{HeatFlatOhmic}). In contrast, $\mathcal{J}_{L}^\texttt{LME}$ in Fig.~\ref{HeatFlatOhmic} remains nonzero in the DSC regime because the RWA was applied to $H_S^{}$, and in this case, the ground state contains no virtual photons. This observation supports our understanding that the thermodynamic decoupling is linked to the proliferation of virtual photons.}

\textcolor{black}{
The experimental detection of the predicted ground-state virtual photons remains an ongoing challenge awaiting demonstration, despite more than a decade of intensive theoretical research~\cite{KonoOptics25}. Typically, many current proposals rely on the nonadiabatic modulation of the system parameters~\cite{ReviewNori}, such as the coupling strength $g$ or the system frequencies $\omega_{c,b}$. Interestingly, there is a recent proposal based on unconventional superconducting quantum circuits with highly efficient conversion of virtual photons into real ones~\cite{GiannelliPPR24}. We believe the circuit-QED architecture is the most promising platform for testing the heat current suppression in the DSC regime, owing to the rapid evolution of their quantum heat transport measurements~\cite{Pekola_RMP_2021} in the strong-coupling regime~\cite{Pekola24arXv}. While current terahertz experiments hold records for large couplings, they have been limited to equilibrium measurements.}



\section{Conclusions}\label{sec:conc}

Within the standard theory of open quantum systems, we have obtained a thermodynamically consistent global (microscopic) master equation in Lindblad form that is valid in the DSC regime, see Eq.~(\ref{rho}). In this regime, the effective light-matter decoupling takes place. The open Hopfield model we used resembled an atomic junction connecting two thermal baths, allowing heat to flow from the hot bath to the cold bath. From our master equation, we were able to identify the temperature-independent dissipation rates of the upper and lower polaritons and, remarkably, to reproduce the breakdown of the Purcell effect using straightforward expressions [see Eqs.~(\ref{decayrates})]. For the thermal transport analysis, we found that the heat current, a nonlocal observable, always vanishes in the DSC regime, a physical process we refer to as thermodynamic decoupling. Using the compact approximate expression we obtained for the population of virtual photons in the ground state, we showed that it grows linearly with $g$ in the DSC, while the heat current decreases as $g^{-1}$. The ability to control the light-matter coupling entails controlling heat transport in ongoing quantum thermodynamics experiments. We hope our results will contribute to this progress.

\section*{Acknowledgments}
R.R.-A. thanks DGAPA-UNAM, Mexico for support under Project No. IA104624. S.P., M.S.-M., and M.S.-G. would like to express their gratitude to SECIHTI, Mexico for their PhD Scholarship.

\appendix

\section{Master equation}\label{Appendix_1}

The master equation used in the main text is 
\begin{equation}
    \dot\rho=-i[H_{\rm diag}^{},\rho]+\mathcal{L}_{\!L}^{}\rho+\mathcal{L}_{\!R}^{}\rho,
\end{equation}
where 
\begin{equation}
    {H}_{\rm diag}^{}=\omega_x^{}\big(A_L^\dagger A_L^{}+{1}/{2}\big)+\omega_y^{}\big(A_R^\dagger A_R^{}+{1}/{2}\big),
\end{equation}
and
\begin{align}\label{L_Lambda}
\mathcal{L}_{\!L}^{}\rho &\equiv
    \gamma_{_L}^{}(\omega_x^{})(f_1^{}+f_2^{})^2\Big\{\big[n(\omega_x,T_L^{})+1\big]\mathcal{L}[A_L^{}]\rho+n(\omega_x^{},T_L^{})\mathcal{L}[A_L^\dagger]\rho\Big\}
    \nonumber\\
    &\qquad+\gamma_{_L}^{}(\omega_y^{})(f_3^{}+f_4^{})^2\Big\{\big[n(\omega_y,T_L^{})+1\big]\mathcal{L}[A_R^{}]\rho+n(\omega_y^{},T_L^{})\mathcal{L}[A_R^\dagger]\rho\Big\},
    \\
\mathcal{L}_{\!R}^{}\rho &\equiv
    \gamma_{_R}^{}(\omega_x^{})(f_5^{}-f_6^{})^2\Big\{\big[n(\omega_x,T_R^{})+1\big]\mathcal{L}[A_L^{}]\rho+n(\omega_x^{},T_R^{})\mathcal{L}[A_L^\dagger]\rho\Big\}
    \nonumber\\
    &\qquad+\gamma_{_R}^{}(\omega_y^{})(f_7^{}-f_8^{})^2\Big\{\big[n(\omega_y,T_R^{})+1\big]\mathcal{L}[A_R^{}]\rho+n(\omega_y^{},T_R^{})\mathcal{L}[A_R^\dagger]\rho\Big\}.
\end{align}
The heat current from the left bath is $\mathcal{J}_L^{}={\rm tr}\{H_{\rm diag}^{}\,\mathcal{L}_{\!L}^{}\rho\}$, which yields
\begin{align}\label{leftheat}
    \mathcal{J}_{\!L}^{}=&
    2\omega_{\!x}^{}\gamma_{_L}(\omega_{\!x}^{})(f_1^{}+f_2^{})^2\big[n(\omega_{\!x}^{},T_L^{})-\langle A_L^\dagger A_L^{}\rangle\big]
    \nonumber\\
    &\qquad+
    2\omega_{\!y}^{}\gamma_{_L}(\omega_{\!y}^{})(f_3^{}+f_4^{})^2\big[n(\omega_{\!y}^{},T_L^{})-\langle A_R^\dagger A_R^{}\rangle\big].
\end{align}
From the time-dependent solutions~(\ref{time_solutions}), at the steady state ($t\rightarrow\infty$) we get $\langle A_L^\dagger A_L^{} \rangle_{\rm SS}^{}=\alpha_2^{}(\alpha_1^{}\!-\!\alpha_2^{})^{-1}$, and $\langle A_R^\dagger A_R^{} \rangle_{\rm SS}^{}=\beta_2^{}(\beta_1^{}\!-\!\beta_2^{})^{-1}$, which results in $\mathcal{J}_L^{\rm SS}$ shown in Eq.~(\ref{HeatCurrent}).

\newcommand{\newblock}{}
\bibliographystyle{unsrt}

\end{document}